\documentclass[runningheads]{llncs}

\usepackage{authblk}
\usepackage{placeins}
\usepackage{graphicx}
\usepackage[colorlinks]{hyperref}
\usepackage{textcomp}
\usepackage{xspace}
\usepackage{caption}
\usepackage{subcaption}
\usepackage{amsmath}
\usepackage{booktabs}
\newcommand{\ModelName}{NeoUNet\xspace}
\newcommand{\DatasetName}{NeoPolyp\xspace}
\newcommand{\CleanDatasetName}{NeoPolyp-Clean\xspace}
\newcommand*{\divby}{\mathrel{\rotatebox{90}{$\hskip-1pt.{}.{}.$}}}%

\begin{document}

	\title{\ModelName{}: Towards accurate colon polyp segmentation and neoplasm detection}
	\titlerunning{Towards accurate colon polyp segmentation and neoplasm detection}

	 \author{
	 	Phan Ngoc Lan\inst{1}
	 	\and Nguyen Sy An\inst{1}
	 	\and Dao Viet Hang\inst{2,3}
	 	\and Dao Van Long\inst{2,3}
	 	\and Tran Quang Trung\inst{4}
	 	\and Nguyen Thi Thuy\inst{5}
	 	\and Dinh Viet Sang\inst{1}
	 }

	 \authorrunning{Phan et al.}

	 \institute{
	 	Hanoi University of Science and Technology, Hanoi, Vietnam
	 	\and Hanoi Medical University, Hanoi, Vietnam
	 	\and The Institute of Gastroenterology and Hepatology, Hanoi, Vietnam
	 	\and University of Medicine and Pharmacy, Hue University, Hue, Vietnam
	 	\and Faculty of Information Technology, Vietnam National University of Agriculture, Hanoi, Vietnam
	 }

	\maketitle

	\begin{abstract}
		Automatic polyp segmentation has proven to be immensely helpful for endoscopy procedures, reducing the missing rate of adenoma detection for endoscopists while increasing efficiency. However, classifying a polyp as being neoplasm or not and segmenting it at the pixel level is still a challenging task for doctors to perform in a limited time. In this work, we propose a fine-grained formulation for the polyp segmentation problem. Our formulation aims to not only segment polyp regions, but also identify those at high risk of malignancy with high accuracy. In addition, we present a UNet-based neural network architecture called \ModelName, along with a hybrid loss function to solve this problem. Experiments show highly competitive results for \ModelName on our benchmark dataset compared to existing polyp segmentation models.

	\end{abstract}

	\section{Introduction}
	Colonoscopy is considered as the most effective procedure for colorectal polyp detection and removal \cite{issa2017colorectal}. Among many histopathological types of precancerous polyps, adenomas with high-grade dysplasia carry the highest risks of developing into colorectal cancer (CRC) \cite{gschwantler2002high}. With over 640,000 deaths each year \cite{bernal2017comparative}, CRC is among the most common types of cancer. As such, the importance of performing colonoscopies to detect and remove high-risk polyps is undeniable.

	However, a review by Leufkens et al. \cite{leufkens2012factors} showed that $20-47\%$ of polyps might have been missed during colonoscopies. There are several factors contributing to this situation, including overloading healthcare systems with an overwhelming number of cases per day and reducing withdrawal time, low-quality endoscopy equipment, or personnel's lack of experience \cite{armin2015visibility,lee2008adequate}. Technologies such as image-enhanced endoscopies and novel accessories were invented and applied to solve this problem. However, cost-effectiveness is still a barrier, especially for limited-resource settings, while performance is not yet ideal. Thus, computer-aided systems have a lot of potentials to improve colonoscopy quality. Recent years have seen very active research in this domain, mostly focused on automatic polyp segmentation and/or detection. Several such works have achieved very high accuracy on benchmark datasets \cite{fan2020pranet,huang2021hardnet,tang2019cu}.

	While automatic segmentation can immensely improve endoscopic performance, it leaves out the difficult task of determining whether a polyp is neoplastic. Neoplastic polyps are precursor lesions to CRC. They require different approaches such as conventional polypectomies, endoscopic mucosal resection, endoscopic submucosal dissection, biopsy, marking, staging for further management such as surgery, or neoadjuvant chemo-radiotherapy. Non-neoplastic polyps, on the other hand, can be removed or left with/without following up during colonoscopies. Classifying neoplastic polyps could be very challenging, especially when the withdrawal time is under overwhelming pressure. The task typically requires well-trained endoscopists with many years of experience, who may still be unable to recognize and characterize some lesions due to tiredness. The critical nature of this task calls for a more fine-grained approach to segment polyps as well as to classify the lesions according to the risk of neoplasm with relative confidence. Several endoscopic systems such as CADEYE (Fujifilm), EndoBrain (Olympus) have integrated this functionality. However, these systems also require magnification and image-enhanced functions, making them costly and unapproachable in most developing countries.

	Polyp segmentation has seen a number of approaches over the years. Traditional machine learning methods based on hand-crafted features \cite{iwahori2013automatic,silva2014toward} rely on color, shape, texture, etc\dots to separate polyps from the surrounding mucosa. These approaches are generally limited, as polyps have very high intra-class diversity and low inter-class variation. The state-of-the-art methods for polyp segmentation in recent years have been deep neural networks. These networks can learn highly abstract and complex image representations to accurately find polyp areas. U-Net \cite{ronneberger2015u} and related models are among the most successful and widely used, all of which feature an encoder-decoder architecture allowing the combination of low-level concrete features and high-level abstract features. Variants such as UNet++ \cite{zhou2019unet++} and ResUNet++ \cite{jha2019resunet++} improved on the original U-Net by adopting a nested architecture, while others (eg. DoubleUNet \cite{jha2020doubleu}) went with a stacking approach. More general techniques in deep learning such as attention and deep supervision have also been incorporated in UNet, and have yielded promising results.

	Fine-grained classification seeks to identify complex \textit{subclasses} instead of simple coarse-grained classes. For example, an image classified as "Dog" may have more fine-grained classes such as "Golden Retriever" or "Pomeranian". While these problems may seem to be similar to multi-class classification, their high inter-class similarity can easily confuse the learning models, leading to low performance.

	In this paper, we first restate the polyp segmentation problem by expanding it with a fine-grained classification aspect. We then propose a UNet-based network architecture to solve it. Specifically, our contributions are:
	\begin{itemize}
		\item To formally describe the polyp segmentation and neoplasm detection problem;
		\item To propose \ModelName{}, a deep neural network architecture designed for the stated problem;
		\item To describe a new dataset, called  \DatasetName, our benchmark dataset for polyp segmentation and neoplasm detection;
		\item To present experimental results for \ModelName on the \DatasetName dataset, including comparisons with existing models for polyp segmentation.
	\end{itemize}

	The rest of the paper is organized as follows. We provide a brief review of related works in Section \ref{sec:related}. Section \ref{sec:problem} describes the polyp segmentation and neoplasm detection problem in detail. The proposed \ModelName is presented in Section \ref{sec:model}. Section \ref{sec:experiment} showcases our experimental studies. Finally, we conclude the paper and highlight future works in Section \ref{sec:conclude}.

	\section{Related work}
	\label{sec:related}
	Convolutional neural networks (CNNs) have claimed state-of-the-art performance in almost every computer vision task in the last few years. Some of the earliest breakout models were AlexNet \cite{krizhevsky2012imagenet} and VGG \cite{simonyan2014very}. These early architectures were still quite limited and suffered from degradation when increasing network depth. ResNet \cite{he2016deep} introduced skip connections that helped smooth out the loss landscape and combat gradient vanishing in very deep networks. GoogLeNet \cite{szegedy2015going} proposed a meticulously-designed multi-branch network with good performance. ResNeXt \cite{xie2017aggregated} also took up a multi-branch approach and applied it to ResNet. EfficientNet \cite{tan2019efficientnet} is a family of networks designed using neural architecture search techniques that provides a range of tradeoffs between accuracy and latency. HarDNet \cite{chao2019hardnet} focused on reducing inference latency by reducing memory traffic.

	Semantic segmentation and segmentation for medical images, in particular, have seen a lot of interest in recent years. Long et al. \cite{long2015fully} adopted several well-known architectures for segmentation using transfer learning techniques. DeepLabV3 \cite{chen2017rethinking} proposed the use of atrous convolutions for dense feature extraction with positive results. PraNet \cite{fan2020pranet} enhanced an FCN-like model with parallel partial decoder and reverse attention. HarDNet-MSEG \cite{huang2021hardnet} adopts the HarDNet backbone as the encoder in an encoder-decoder structure. The network achieved state-of-the-art performance on the Kvasir-SEG dataset while also improving inference latency.

	U-Net \cite{ronneberger2015u} was one of the first successful CNNs applied in medical imaging. The architecture features an encoder-decoder design, combining low-level features on the encoder branch with high-level features on the decoder branch. Numerous works have proposed improvements to the original UNet. UNet++ \cite{zhou2019unet++} and ResUNet++ \cite{jha2019resunet++} used a nested architecture, with multiple levels of cross-connections. DoubleUNet \cite{jha2020doubleu} stacked two UNets sequentially, using VGG-16 as the encoder backbone, with squeeze and excitation units and ASPP modules. While outperforming previous methods on several datasets, DoubleUNet is limited in terms of information flow between the two UNets. Tang et al. \cite{tang2019cu} tackled this limitation with Coupled U-Net (CUNet), which adds skip connections between UNet blocks. Attention-UNet \cite{oktay2018attention} introduced attention gates which filters for useful salient features. Abraham et al. \cite{abraham2019novel} proposed a novel loss function in conjunction with the Attention-UNet architecture, combined with multi-scale input and deep supervision.


	Due to the heterogeneous quality of different endoscopic systems, the lack of public datasets for polyp characterization is a challenge. Among the most similar research to ours is that of Ribeiro et al. \cite{ribeiro2016exploring}, who presented several approaches - including CNNs and hand-crafted features - for polyp classification. The authors extracted a dataset of 100 polyp images from endoscopy videos, each containing exactly one polyp. Several CNN models were tested on this dataset, including VGG, AlexNet, GoogLeNet, etc\dots\ A primary drawback for this approach is that classification has to be done after detection or segmentation. In other words, the problem is approached in two stages. While it is possible to combine the polyp detection modules with classification, this method can be inefficient and cumbersome, especially for systems with real-time requirements or running on embedded devices. In this paper, we present an end-to-end model for both segmentation and classification of polyps to overcome these limitations.

	\section{Polyp segmentation and neoplasm detection}
	\label{sec:problem}
	The problem presented in this work is an expansion of polyp segmentation, focusing on more fine-grained classification. In polyp segmentation, given an input image, we need to output a binary mask where each pixel's value is either 1 (the pixel is part of a polyp) or 0 (the pixel is part of the background).

\begin{figure}[tbh]
		\centering
		\subcaptionbox{Input image}{
			\includegraphics[width=0.27\textwidth]{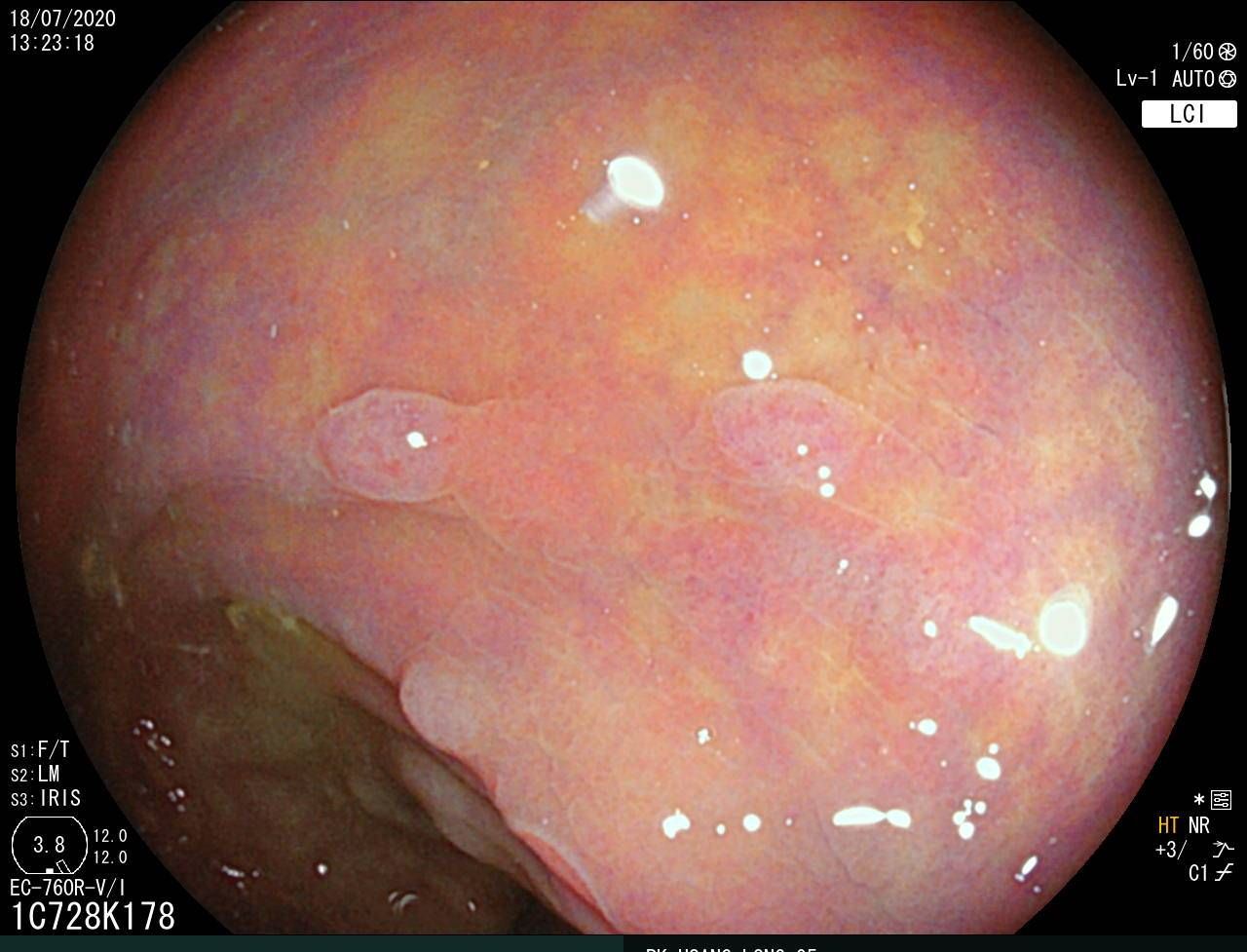}
		}
		\subcaptionbox{Polyp segmentation}{
			\includegraphics[width=0.27\textwidth]{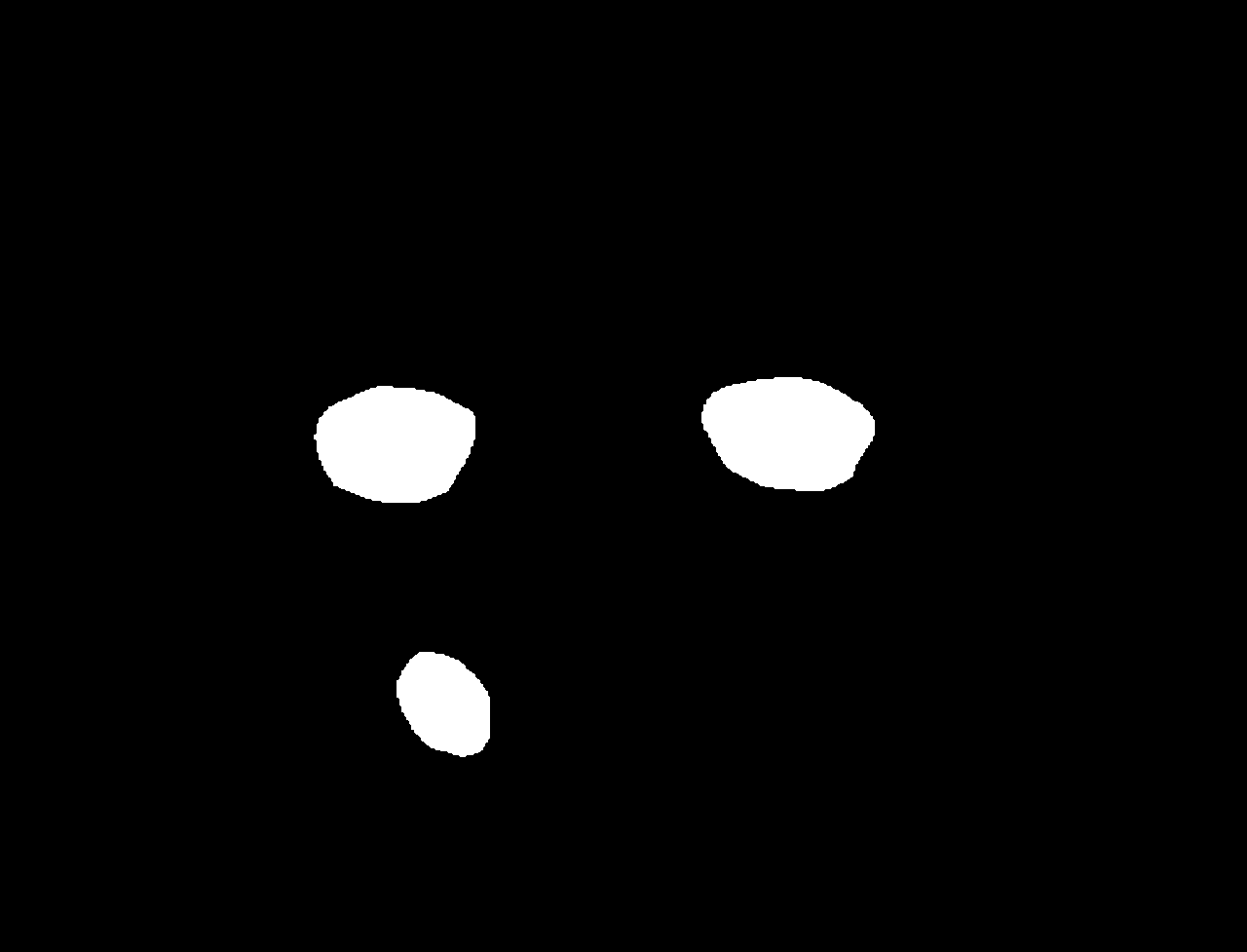}
		}
		\subcaptionbox{PSND}{
			\includegraphics[width=0.27\textwidth]{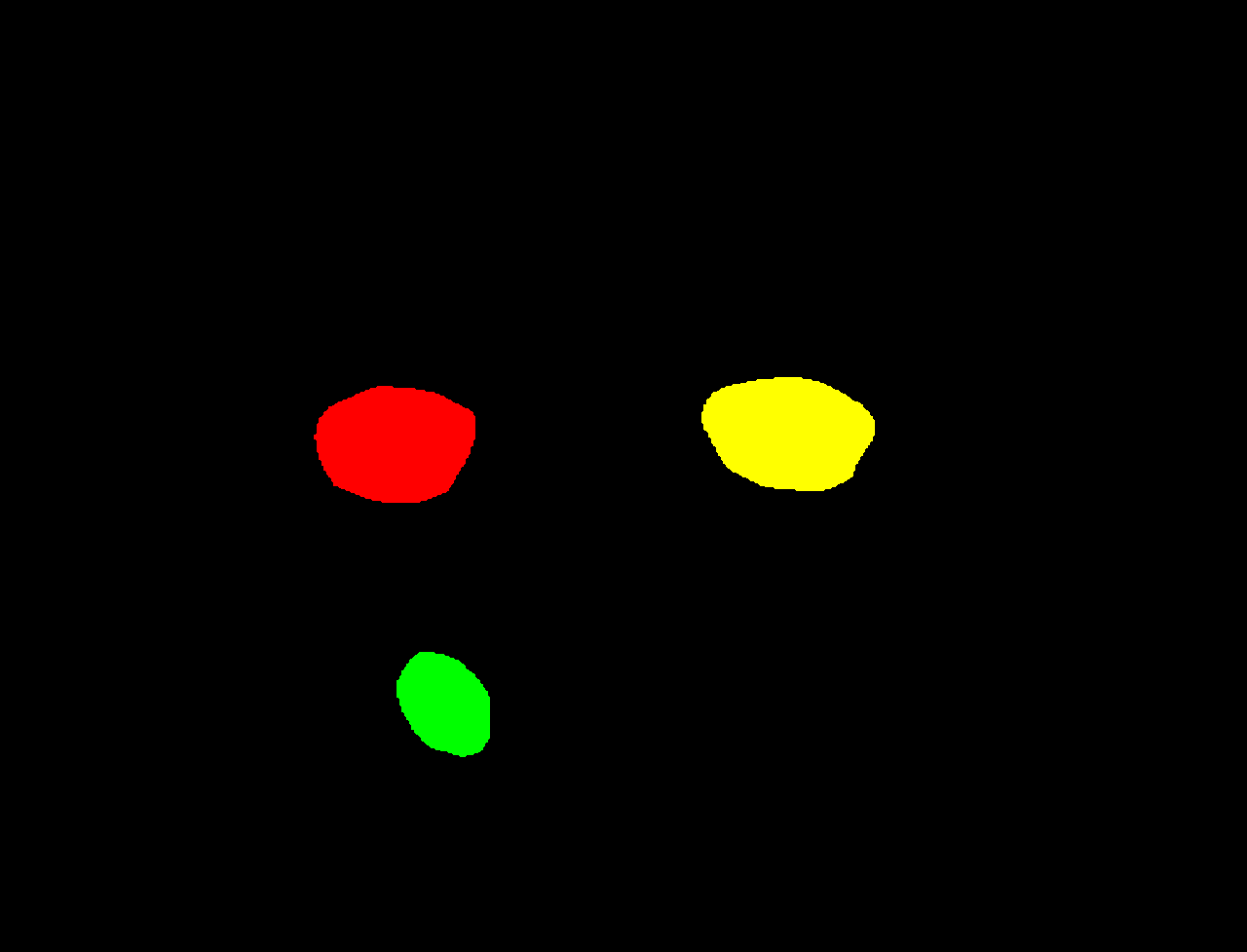}
		}
		\caption{Expected outputs for polyp segmentation and PSND. Black regions denote background pixels. White regions denote polyp regions. Green, red and yellow regions denote non-neoplastic, neoplastic and unknown polyp regions, respectively.}
		\label{fig:example}
	\end{figure}

	The polyp segmentation and neoplasm detection problem (PSND) expects each pixel in the segmentation mask to have one of four values (see Fig. \ref{fig:example} for examples):
	\begin{itemize}
		\item 0 if the pixel is part of the image background;
		\item 1 if the pixel is part of a non-neoplastic polyp;
		\item 2 if the pixel is part of a neoplastic polyp;
		\item 3 if the pixel is part of a polyp with unknown neoplasticity.
	\end{itemize}

	PSND presents several unique challenges. First, the surface pattern of a polyp could be homogenous or heterogeneous, with different areas of texture which requires experienced endoscopists to evaluate carefully. In one lesion, a neoplastic area may only take up a portion of the polyp's surface, causing further difficulties for machine learning models. This property is the primary distinction between our problem and generic multi-class segmentation. While datasets such as PASCAL-VOC have many more classes, they are primarily well-defined and highly distinguishable (e.g., car, bird, person, etc\dots). Moreover, from the medical perspective, a misclassification could lead to biased decisions. False positives could result in over indications of endoscopic interventions or surgery, while false negatives could result in delaying suitable treatment.

	Polyps considered to have "unknown" neoplasticity also pose specific challenges to machine learning models. We generally do not want automatic systems to learn and output "unknown" pixels. Instead, it is much more beneficial to produce a classification between neoplastic and non-neoplastic, as this provides more insight to physicians while also reducing inter-class similarity. At the same time, "unknown" polyps are still helpful for learning how to segment without classification.

	The challenges mentioned above serve as motivation for the proposed \ModelName model, which we shall describe in detail in the next section.

	\section{\ModelName}
	\label{sec:model}
	\subsection{Architecture}
	\ModelName is a U-Net architecture similar to the one introduced in \cite{abraham2019novel}, with several key differences. The encoder backbone uses the HarDNet68 architecture \cite{chao2019hardnet}, comprising of Harmonic Dense blocks (HDB). Outputs from each encoder level are passed to corresponding decoder blocks through attention gate modules. All decoder blocks also have output layers producing multi-class segmentation masks at their corresponding scale level. These output layers allow us to train the network using deep supervision, which boosts the network's stability and convergence rate.

	To take advantage of public ImageNet-trained HarDNet models, we do not make modifications to the backbone structure and thus cannot use joint multi-scale inputs. However, the benefits of pretraining can outweigh this limitation.

\begin{figure}
		\centering
		\includegraphics[width=\textwidth]{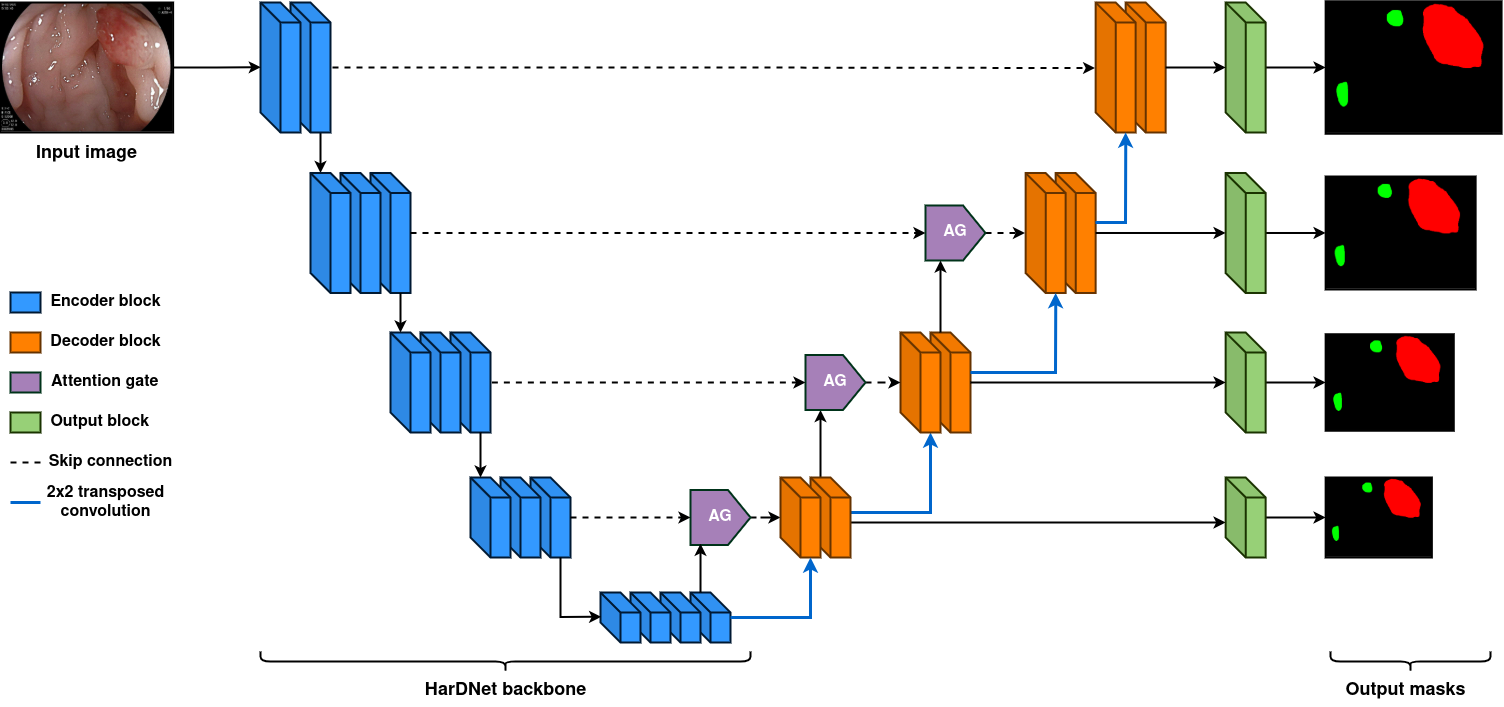}
		\caption{Overview of \ModelName{}'s architecture}
		\label{fig:model-overview}
	\end{figure}

	The architecture's overview is shown in Fig. \ref{fig:model-overview}.

	\subsubsection{Encoder backbone}
	HarDNet (Harmonic DenseNet) \cite{chao2019hardnet} is a CNN architecture inspired by DenseNet \cite{huang2017densely}. DenseNet's core principle is encouraging feature reuse through skip connections: each layer in a Dense Block receives the concatenated feature map of every preceding layer, essentially having a \textit{global state} through which high-value features can be shared. While achieving high accuracy, DenseNet also has a large memory footprint, leading to low throughput and increased latency due to memory traffic.

\begin{figure}
		\centering
		\includegraphics[width=0.95\textwidth]{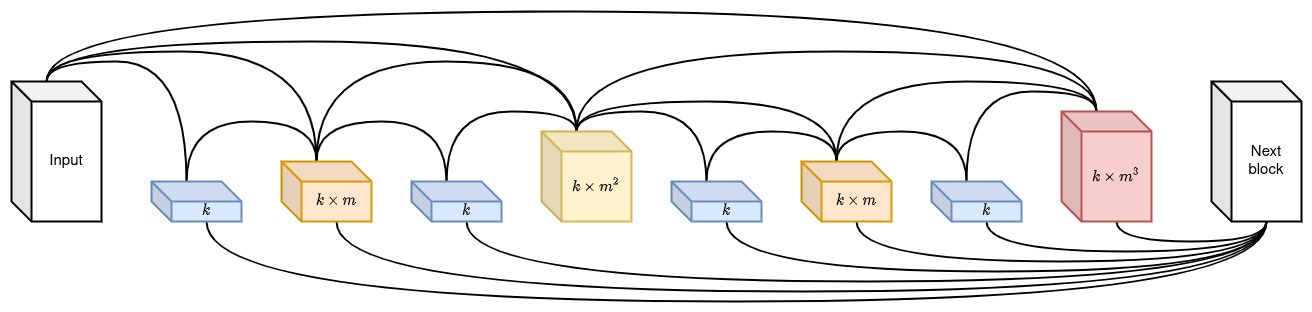}
		\caption{Structure of an example Harmonic Dense Block. The value on each layer denotes the number of output channels.}
		\label{fig:model-hdb}
	\end{figure}

	To create a better tradeoff between accuracy and memory, HarDNet sparsifies DenseNet by reducing the number of skip connections. Specifically, a layer $k$ in a HarD Block (HDB) receives a feature map from layer $k - 2^n$ if $2^n$ divides $k$ ($n \geq 0$, $k - 2^n \geq 0$). In addition, layers whose indices are divisible to large powers of 2 are more "influential", as their feature maps are more frequently reused. Such layers are given more convolutional kernels (more output channels). Concretely, the number of output channels for layer $l$ with a growth rate $k$ is $k \times m^n$, where $n = max \{\nu \ | \ l \divby 2^{\nu} \}$. $m$ can be considered as the compression factor, where $m = 2$ means all layers have the same number of input feature maps. Fig. \ref{fig:model-hdb} illustrates the structure of a HDB.

	HarDNet68 further changes DenseNet architecture by removing the global dense connections, using Max Pooling for downsampling, and having dedicated growth rates for each HDB. HarDNet68 has 5 "strides" that reduce the feature map size by 2, 4, 8, 16, and 32 times compared to the original image. When incorporated in \ModelName as the encoder backbone, each stride serves as an encoder block.

	\subsubsection{Attention Gate}
\begin{figure}
		\centering
		\includegraphics[width=0.9\textwidth]{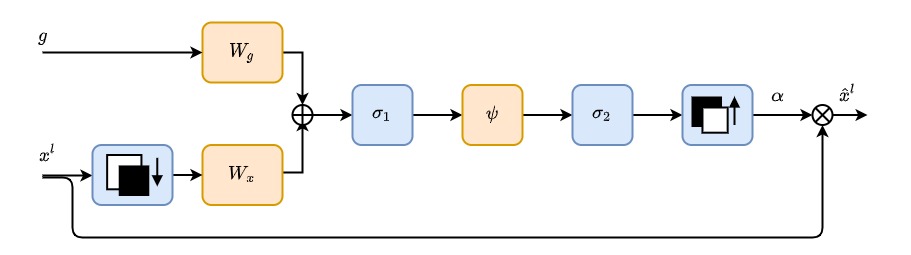}
		\caption{Diagram of the additive attention gate module \cite{oktay2018attention}}
		\label{fig:attn}
	\end{figure}

	The information carried in skip connections between encoder and decoder blocks can generally be noisy, as finer feature maps contain more local features that may be irrelevant. Thus, we use additive attention gates \cite{oktay2018attention} on all such connections, except for the top-most blocks. An attention gate filters the input tensor using a set of attention coefficients $\alpha_i$, calculated from the input $x^l$ and the gating signal $g$ (see Eq. \eqref{eq:attn-1} and Eq. \eqref{eq:attn-2} \cite{oktay2018attention} for details).

	\begin{align}
	\label{eq:attn-1}
	q^l_{att} & = \psi^T (\sigma_1 (W^T_x x^l_i + W^T_g g_i + b_g)) + b_{\psi} \\
	\label{eq:attn-2}
	\alpha^l_i & = \sigma_2 (q^l_{att}(x^l_i, g_i; \Theta_{att}))
	\end{align}
	where $\sigma_1$ denotes the ReLU function, $\sigma_2$ denotes the sigmoid function. $\Theta_{att} = (W_x, W_g, b_g, \psi, b_{\psi})$ is the set of learnable parameters. The attention gate's output is the filtered tensor $\hat{x}^l = x^l . \alpha$. Fig. \ref{fig:attn} illustrates the attention gate's structure.

	\subsubsection{Decoder and output block}
	Each decoder block consists of 2 sequential sets of \{Convolution, Batch Norm, Leaky ReLU\} layers. The output of the previous decoder block is concatenated with the attention gate output (carrying filtered information from the encoder) to form the current decoder's input.

	Each output block is a simple 1x1 convolution layer with two output channels, followed by the sigmoid activation. Each channel corresponds to the output mask for one class. Segmentation maps without classification can be inferred using the element-wise OR operator.

	During training, we upsample output masks from all levels to the original image size and compute the loss values for each mask. These values are then summed to form the final loss for backpropagation. This form of deep supervision forces different levels of abstract features to make sense semantically, allowing faster convergence and improving stability.

	\subsection{Loss function}
	The loss function for training \ModelName is a weighted sum of a segmentation loss and a multi-class loss, as shown in Eq. \eqref{eq:L}.
	\begin{equation}
	\label{eq:L}
	\mathcal{L}(M_{ct}, M_{st}, M_{cp}, M_{sp}) = w_c \mathcal{L}_{c}(M_{ct}, M_{cp}) + w_s \mathcal{L}_{s}(M_{st}, M_{sp})
	\end{equation}

	The multi-class loss $\mathcal{L}_{c}$ reflects the difference between the ground truth multi-class mask $M_{ct}$ and the predicted multi-class mask $M_{cp}$. $\mathcal{L}_{c}$ is averaged from Binary Cross Entropy loss and Focal Tversky loss \cite{abraham2019novel} (see Eq. \eqref{eq:Lc}). This is the primary loss function that drives the model toward making accurate class-specific segmentation.

	\begin{equation}
	\label{eq:Lc}
	\mathcal{L}_{c}(M_{ct}, M_{cp}) = \frac{BCE(M_{ct}, M_{cp}) + FocalTversky(M_{ct}, M_{cp})}{2}
	\end{equation}

	Tversky loss \cite{salehi2017tversky} can be seen as a generalization of Dice loss, with two hyperparameters $\alpha$ and $\beta$ that control the effect of false positives and false negatives. Focal Tversky loss \cite{abraham2019novel} extends Tversky loss by combining ideas from Focal loss \cite{lin2017focal}, which adds a parameter $\gamma$ that helps the model to focus on hard examples. For \ModelName, we set $\alpha = 1 - \beta = 0.3$ for Tversky loss, meaning the model prioritizes recall over precision (since smaller polyps are likely to be missed), and $\gamma = \frac{4}{3}$ to focus slightly more on hard examples.

	The segmentation loss $\mathcal{L}_{s}$ is calculated using the ground truth binary segmentation mask $M_{st}$, and the predicted binary segmentation mask $M_{sp}$. $M_{sp}$ can be inferred from the multi-class mask $M_{cp}$ as in Eq. \eqref{eq:Msp}.

	\begin{equation}
	\label{eq:Msp}
	M_{sp}(i, j) = \begin{cases}
	0 \text{ if } M_{cp}(i, j) = 0 \\
	1 \text{ otherwise }
	\end{cases}
	\end{equation}

	$\mathcal{L}_{s}$ is averaged from Binary Cross Entropy loss and Tversky loss (see Eq. \eqref{eq:Ls}). This is a secondary loss function that ensures the model maintains high segmentation accuracy. In addition, the segmentation loss allows \ModelName to take advantage of training data marked as "Unknown."

	\begin{equation}
	\label{eq:Ls}
	\mathcal{L}_{s}(M_{st}, M_{sp}) = \frac{BCE(M_{st}, M_{sp}) + Tversky(M_{st}, M_{sp})}{2}
	\end{equation}

	The parameters $w_c$ and $w_s$ controls the level of effect each loss has on the training process. We find that \ModelName performs best when $w_c = 0.75$ and $w_s = 0.25$.

	During training, pixels that are part of "unknown" polyps only contribute to the segmentation loss and are ignored by the multi-class loss. This mechanism allows the model to freely label "unknown" polyps based on existing features while still benefitting from partially labeled data.

	\section{Experiments and discussion}
	\label{sec:experiment}
	\subsection{Benchmark dataset}
	\label{sec:experiemn}
	A dataset of 7,466 annotated endoscopic images is curated in order to train and benchmark the proposed \ModelName{}. These images are captured directly during endoscopic recording, including all four lighting modes: WLI (White Light Imaging), FICE (Flexible spectral Imaging Color Enhancement), BLI (Blue Light Imaging), and LCI (Linked Color Imaging). The dataset also includes polyps in all Paris classifications \cite{lambert2003paris} \footnote{Labels for this type of classification are not available} (Ip, Is, IIa, IIb, IIc, III), as well as images without any polyps. The patient's identifying information is removed from each image to ensure anonymity.

	Annotations (including segmentation and classification) are added to each image independently by two experienced endoscopists. Matching annotation labels are accepted into the dataset, while those without full consensus from annotators or declared unknown by at least one annotator are marked with the label "Unknown". The dataset is randomly split into a training set of 5,966 images and a test set of 1,500 images.

	In order to compare \ModelName with existing approaches, we also create a filtered dataset without any "Unknown" labels. This dataset consists of 5,277 training images and 1,353 test images. We denote the full dataset as \DatasetName and the filtered version as \CleanDatasetName{}.

\begin{figure}[]
		\centering
		\includegraphics[width=0.45\textwidth]{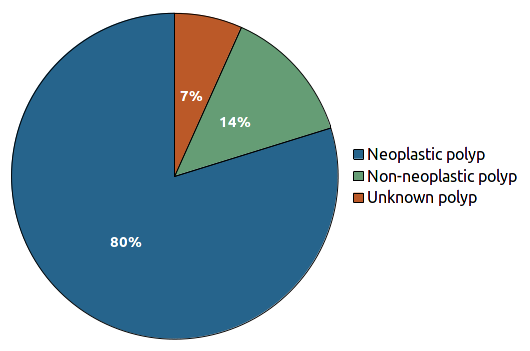}
		\caption{Pixel-wise distribution of polyp class labels in the \DatasetName dataset. Percentages are calculated on polyp pixels only (not including background pixels).}
		\label{fig:data_dist}
	\end{figure}

	Due to their diverse nature, neoplastic polyps take up a majority of the polyps present in \DatasetName (see Fig. \ref{fig:data_dist}). This data imbalance, combined with the inherent challenges of PSND, creates a difficult benchmark for models to overcome.

	\subsection{Experiment setup}
	We perform two experiments to validate the effectiveness of our proposed \ModelName:
	\begin{itemize}
		\item A comparison with baseline models: we compare \ModelName with U-Net \cite{ronneberger2015u}, PraNet \cite{fan2020pranet}, and HarDNet-MSEG \cite{huang2021hardnet} on the PSND task;
		\item A comparison on \DatasetName and \CleanDatasetName{}: we compare the performance of \ModelName when trained on \DatasetName and \CleanDatasetName{}, in order to show the benefits of learning on "Unknown"-labeled data.
	\end{itemize}

	In both experiments,common used metrics are employed for performance measurement:
	\begin{itemize}
		\item Polyp segmentation performance (without classification), measured using Dice score and IoU score. We denote these metrics as $\text{Dice}_{seg}$ and $\text{IoU}_{seg}$;
		\item Segmentation performance on each class (non-neoplastic and neoplastic), measured using Dice score and IoU score. We denote these metrics as $\text{Dice}_{non}$, $\text{IoU}_{non}$, $\text{Dice}_{neo}$ and $\text{IoU}_{neo}$, respectively.
		\item Inference speed for each model, measured as frames per second (FPS). This metric is only of significance in the first experiment.
	\end{itemize}

	Dice and IoU are calculated pixel-wise on the entire test set (micro-averaged, not averaged over each image). Evaluation is done on images resized to $352 \times 352$. To measure inference speed for a model, we average the latency of inferring 100 test images with a batch size of 1.

	Additionally, the following settings and techniques are applied to the training process in both experiments:
	\begin{itemize}
		\item To avoid bias due to class imbalance in the dataset, we oversample images containing non-neoplastic polyps during training such that $P_{non} \approx P_{neo}$, where $P_{non}$ and $P_{neo}$ are the number of pixels containing non-neoplastic and neoplastic polyps, respectively;
		\item \ModelName is trained using Stochastic Gradient Descent with Nesterov momentum and a learning rate of $0.001$. The learning rate $lr$ is adjusted according to a combination of warmup and cosine annealing schedule. Concretely, $lr$ increases linearly up to the target value in the first $t_w$ epochs. In the remaining epochs, $lr$ is annealed following the cosine function \cite{ilya2016sgdr}.
		\item Each training batch is put through the network in 3 different scales: $448 \times 448$, $352 \times 352$, and $256 \times 256$;
		\item Data augmentation is employed to improve models' generality. Specifically, we use five types of augmentation: rotate, horizontal/vertical flip, motion blur, and color jittering. Augmentation is performed on the fly with a probability of 0.7 (i.e., each image has a 70\% chance of being augmented each time it is selected for a training batch).
	\end{itemize}


	\ModelName and the original U-Net are implemented in Python 3.6 using the PyTorch \cite{NEURIPS2019_9015} framework. We use the official PraNet \footnote{https://github.com/DengPingFan/PraNet} and HarDNet-MSEG\footnote{https://github.com/james128333/HarDNet-MSEG} implementations to perform comparison. All training is done on a machine with a 3.7GHz AMD Ryzen 3970X CPU, 128GB RAM, and an NVIDIA GeForce GTX 3090 GPU. We use a Google Colab instance with 2 CPU cores and an NVIDIA Tesla V100 GPU to measure inference speed.

	\subsubsection{Comparison with the baseline models}
	This experiment uses the \CleanDatasetName dataset, as the baseline models do not handle "Unknown" labels in the input data. Each model is first pretrained for 200 epochs on the polyp segmentation training data from \cite{jha2019resunet++} (including images from the Kvasir-SEG and CVC-ClinicDB dataset). We then train each model until convergence on the \CleanDatasetName dataset (by swapping the final output layers with 2-channel convolutional blocks).

	\subsubsection{Comparison on \DatasetName and \CleanDatasetName}
	For this experiment, \ModelName is first trained on the polyp segmentation task using data from \cite{jha2019resunet++}. We then compare the performance when training this model to convergence on the \DatasetName and \CleanDatasetName datasets.

	\subsection{Results and discussion}

	\subsubsection{Comparison with baseline models}
	Table \ref{tab:exp1} shows the performance metrics for \ModelName and HarDNet-MSEG on the \CleanDatasetName dataset. We can see that all four models struggle with the non-neoplastic class. The highest Dice score for this class is only $0.713$ from \ModelName, significantly lower than on the neoplastic class ($0.887$) and the segmentation task ($0.902$). This is likely due to the heavy class imbalance in the training dataset. Although oversampling and Focal loss partially mitigates the issue, there is still a noticeable drop in the performance.

	\ModelName significantly outperforms all baseline models in all Dice and IoU metrics. Against the best baseline, PraNet, our model achieves better scores by $\sim 2\%$ in each metric. The segmentation task sees the most improvement from \ModelName, with a $1.6\%$ increase in Dice score and $2.6\%$ increase in IoU. These results show that the combination of HDB and attention gates used in \ModelName is very effective for the PSND problem. Additionally, the use of the secondary segmentation loss in \ModelName has helped in maintaining segmentation accuracy. Meanwhile, U-Net performs significantly worse than the other three methods.

	In terms of speed, HarDNet-MSEG is the fastest model by a large margin, achieving 77.1 FPS. This is consistent with the results in \cite{huang2021hardnet}, as speed is a major focus for this model. PraNet is the slowest of the four models, at only 55.6 FPS. \ModelName and U-Net have similar speeds at 68.3 and 69.6 FPS, respectively. Overall, despite being slower than HarDNet-MSEG and U-Net, \ModelName provides a good tradeoff between accuracy and speed, while still being faster and more accurate than PraNet.

	\begin{table}[]
		\centering
		\caption{Performance metrics on the \CleanDatasetName test set for U-Net, PraNet, HarDNet-MSEG, and \ModelName}
		\label{tab:exp1}
		\begin{tabular}{@{} l r r r r r r r@{}}
			\toprule

			\multicolumn{1}{c}{Method} & \multicolumn{1}{c}{$\text{Dice}_{seg}$} & \multicolumn{1}{c}{$\text{IoU}_{seg}$} & \multicolumn{1}{c}{$\text{Dice}_{non}$} & \multicolumn{1}{c}{$\text{IoU}_{non}$} & \multicolumn{1}{c}{$\text{Dice}_{neo}$} & \multicolumn{1}{c}{$\text{IoU}_{neo}$} & \multicolumn{1}{c}{FPS} \\ \midrule
			U-Net \cite{ronneberger2015u} & 0.785 & 0.646 & 0.525 & 0.356 & 0.773 & 0.631 & 69.6 \\
			HarDNet-MSEG \cite{huang2021hardnet} & 0.883 & 0.791 & 0.659 & 0.492 & 0.869 & 0.769 & \textbf{77.1} \\
			PraNet \cite{fan2020pranet} & 0.895 & 0.811 & 0.705 & 0.544 & 0.873 & 0.775 & 55.6 \\
			\ModelName & \textbf{0.911} & \textbf{0.837} & \textbf{0.720} & \textbf{0.563} & \textbf{0.889} & \textbf{0.800} & 68.3 \\

			\bottomrule
		\end{tabular}
	\end{table}

	\subsubsection{Comparison on \DatasetName and \CleanDatasetName}
	Table \ref{tab:exp2} shows performance metrics for \ModelName when trained on \DatasetName and \CleanDatasetName. We notice a slight drop in accuracy for the same model when testing on \DatasetName compared to \CleanDatasetName, since images containing "Unknown" polyps are typically more challenging. Interestingly, the use of "Unknown"-label data shows improvement for all metrics, not just segmentation. While the difference is slight ($0.2-0.5\%$), it shows that making use of data for one task can yield benefits in other tasks. In this case, better segmentation masks invariantly improve classification, as these two tasks are intertwined.

	\begin{table}[tbh]
		\centering
		\caption{Performance metrics for \ModelName when training on \DatasetName and \CleanDatasetName, measured on the \DatasetName test set}
		\label{tab:exp2}
		\begin{tabular}{@{} l r r r r r r@{}}
			\toprule
			\multicolumn{1}{c}{Training dataset} & \multicolumn{1}{c}{$\text{Dice}_{seg}$} & \multicolumn{1}{c}{$\text{IoU}_{seg}$} & \multicolumn{1}{c}{$\text{Dice}_{non}$} & \multicolumn{1}{c}{$\text{IoU}_{non}$} & \multicolumn{1}{c}{$\text{Dice}_{neo}$} & \multicolumn{1}{c}{$\text{IoU}_{neo}$} \\ \midrule
			\CleanDatasetName & 0.906 & 0.828 & 0.725 & 0.569 & 0.888 & 0.799 \\
			\DatasetName & 0.908 & 0.831 & 0.729 & 0.573 & 0.891 & 0.804 \\
			\bottomrule
		\end{tabular}

	\end{table}

	\section{Conclusion}
	\label{sec:conclude}
	This paper has presented the polyp segmentation and neoplasm detection problem, a challenging combination of fine-grained classification and semantic segmentation. To solve this problem, we propose \ModelName, a UNet-based architecture incorporating attention gates, an efficient HarDNet backbone, and a hybrid loss function to take advantage of unknown labels. Our experiments show very competitive results when compared to existing models for polyp segmentation.

	We hope that this work can be a basis for further improvements on this challenging problem. Our future works include improving classification accuracy, especially for non-neoplastic polyps.

	 \section{Acknowledgments}
	 This work was funded by Vingroup Innovation Foundation (VINIF) under project code VINIF.2020.DA17. Phan Ngoc Lan was funded by Vingroup Joint Stock Company and supported by the Domestic Master/Ph.D. Scholarship Programme of Vingroup Innovation Foundation (VINIF), Vingroup Big Data Institute (VINBIGDATA), code VINIF.2020.ThS.BK.02.

	\bibliographystyle{splncs04}
	\bibliography{paper}
\end{document}